\documentclass[11pt,a4paper,nofootinbib]{revtex4}
\usepackage[utf8]{inputenc}
\usepackage[T1]{fontenc}
\usepackage{lmodern}
\usepackage[margin=1in]{geometry}
\usepackage{microtype}
\begin{document}
\title{Quantum foundations for quantum technologies in the International Year of Quantum (2025)}
\author{A. Bassi\footnote{Corresponding author.}}
\email{abassi@units.it}
\affiliation{Department of Physics, University of Trieste, Strada Costiera 11, 34151 Trieste, Italy}
\affiliation{INFN, Sezione di Trieste, Via Valerio 2, 34127 Trieste, Italy.}

\begin{abstract}
From the very beginning, Quantum Mechanics has been accompanied by crucial foundational questions: the possibility of visualizing physical processes, the limits of measurement epitomized by Heisenberg’s uncertainty principle, the existence of a deeper underlying reality with additional degrees of freedom, the role of measurements, and the status of locality. Long regarded as philosophical speculations, these issues were progressively reformulated into precise mathematical statements and ultimately subjected to experimental verification. The trajectory proved unpredictable: questions once dismissed as metaphysical gave rise to experimental platforms, which in turn matured into devices and technologies powering quantum computation, communication, and sensing. Yet this development is not unidirectional: advances in technology also feed back into foundations, enabling tests of principles that were previously out of reach—for example, whether quantum superposition persists at larger and larger scales and whether reality, gravity included, is fundamentally quantum. In this way, the dialogue between foundational inquiry and technological progress continues to shape both our theoretical understanding and the practical realization of quantum phenomena.
\end{abstract}

\maketitle

\section{1925-2025: from debates to devices}
Quantum Mechanics has challenged the scientific worldview since its inception, and the famous debate between Einstein and Bohr~\cite{BohrEinstein} made the challenge explicit: for Einstein, Quantum Mechanics was astonishingly accurate yet incomplete—correct to a very high degree but not offering the full picture of elementary processes—whereas for Bohr it was the complete and correct framework for microscopic phenomena and their observation. 

Well before the famous EPR paper, Einstein had already pressed the point\footnote{Don Howard~\cite{Howard} documents threads of Einstein’s thinking about the incompleteness of Quantum Theory even before 1927.} at the 1927 Solvay Conference~\cite{solvay}: he considered a single particle whose wave function is spread out in space and asked what happens when its position is detected. He remarked that “{\it the interpretation according to which $|\psi|^2$ expresses the probability that} this {\it particle is found at the given point assumes an entirely peculiar mechanism of action at a distance, which prevents the wave continously distributed in space from producing an action in} two {\it places on the screen. In my opinion, one can remove this objection only in the following way, that one does not describe the process solely by the Schr\"odinger wave, but at the same time one localizes the particle during the propagation}.” In other words, for a delocalized wave function, a position measurement that (randomly) finds the particle here entails with certainty that it cannot be found elsewhere; then, either there are nonlocal correlations---what happens here constrains what can happen there, in principle far away---or the outcomes were predetermined, as in a classical statistical model. Taking locality for granted, as Einstein did, only the latter option remains: the properties of the particle must be predetermined, and therefore the quantum-mechanical description offered by the wave function is incomplete.

The  1935 EPR paper~\cite{epr} presents essentially the same argument: locality + quantum correlations in far away places $\rightarrow$ pre-determined properties,  which in turn implies the incompleteness of Quantum Theory; as a matter of fact, the paper ends by saying: “{\it While we have thus shown that the wave function does not provide a complete description of the physical reality, we left open the question of whether or not such a description exists. We believe, however, that such a theory is possible.}”  The EPR paper is important because it uses a decisive new ingredient: entanglement between two particles, to encode perfect correlations across space-like separations. That conceptual move furnished the raw material for Bell’s inequalities and for later generations to explore quantum resources to their full potential. 

Bohr’s reply the same year~\cite{Bohr} defended completeness, by insisting on the holistic character of quantum phenomena and on complementarity as a constraint on what can be jointly defined and observed. But actually, the first reaction of Bohr to Einstein's argument at the 1927 Solvey conference was~\cite{solvay} : ``{\it I feel myself in a very difficult position because I don't understand what precisely is the point which Einstein wants to [make].  No doubt it is my fault.}'' The community’s reaction mixed admiration for Einstein  with a tendency to dismiss his concern as metaphysical scruple; as Pauli quipped in a 1954 letter to Born~\cite{Pauli}, {\it “One should no more rack one’s brain about the problem of whether something one cannot know anything about exists all the same than about the ancient question of how many angels are able to sit on the point of a needle. But it seems to me that Einstein’s questions are ultimately always of this kind.”} 

This historical trajectory matters: since the EPR era, entanglement has transformed from a philosophical irritant into the central resource of quantum information science, and today’s quantum technologies owe their superior capabilities to precisely those features—entanglement and coherence—that enable tasks unattainable by classical means. That is an apt message for the {\it International Year of Quantum (2025)}: foundations and technology advance together in a feedback loop in which fundamental questions are distilled into mathematical models susceptible to experimental verification; those models stimulate laboratory tests and generate new technological platforms and protocols, which in turn make it possible to address sharper open questions.

\section{Bell's turn: nonlocality and the rise of quantum networks}

Although the EPR paper initially received comparatively little  attention, Bohm gave it new life and clarity in his textbook reformulation of Quantum Theory~\cite{Bohm}, recasting the argument in terms of a pair of spin-$1/2$ particles prepared in the singlet state, and then sent to distant parties Alice and Bob. Because for a singlet state spin measurements are perfectly anti-correlated along any common axis, if Alice measures the spin along the $z$  direction, represented by the operator $\sigma_{z}$ and obtains, say, $+1$, she can infer with certainty that Bob’s outcome for $\sigma_{z}$ will be $-1$. Under the locality assumption—Alice’s choice and outcome cannot affect Bob’s distant system—this  licenses an “element of reality” for Bob particle’s $\sigma_{z}$ value prior to his measurement. But the same reasoning holds if Alice instead measures along $x$ (or any other direction), so Bob's particle must likewise possess predetermined values for non-commuting observables such as $\sigma_{x}$ and $\sigma_{z}$ simultaneously. The conclusion is again: locality + perfect quantum correlations $\rightarrow$ predetermined properties $\rightarrow$ the wave function is incomplete. Bohm’s formulation preserved the spirit of EPR while stripping away technicalities, recasting the argument in terms of states and observables that can be more easily prepared and measured in the laboratory, thereby setting the stage for Bell’s inequalities.

The unexpected twist in this story was Bell’s 1964 theorem~\cite{Bell1}. Bell shared Einstein’s worries about the status of Quantum Theory as a fundamental description of nature and wrote about them repeatedly~\cite{BellSp}. He was especially intrigued by de Broglie–Bohm pilot-wave theory~\cite{deBroglie1924,Bohm1952,DurrTeufel2009} because it offered a clear, unambiguous dynamics for particles guided by the wave function, thus avoiding any special status for “measurements'' and ``observers''. But the price was explicit nonlocality, and this troubled him. Bell then asked whether such nonlocality was a peculiarity of Bohmian mechanics or an unavoidable feature of \emph{any} theory reproducing the quantum predictions for entangled systems. This is the context in which he derived his famous theorem.

As Bell explains in his 1964 paper, the starting point is precisely the EPR logic in the version of Bohm: under \emph{locality}, the perfect anticorrelations of a singlet pair imply that the outcomes of spin measurements are predetermined (there exist what now are called hidden varaibles). Bell’s next step shows that any \emph{local hidden-variable} theory must satisfy certain constraints—Bell inequalities—on joint statistics for space-like measurements along different directions. Quantum Mechanics violates those constraints for appropriate choices of settings. 

Two remarks are in order. First, in the EPR era the focus was on perfect (anti)correlations obtained when Alice and Bob measure along the {\it same} spin axis. Bell’s key insight was to step away from that trivial case and ask what happens for different measurement directions—precisely where one might expect nothing surprising. The surprise was there: locality imposes quantitative constraints on those cross-axis statistics, captured by Bell’s inequality. Second, “elements of reality” (predetermined values, hidden variables, or what now is sometimes called realism) are not an independent assumption of Bell’s theorem. In Bell’s 1964 analysis they arise only via the EPR step: locality plus perfect (anti)correlations $\rightarrow$ predetermined outcomes. Bell was clear about this~\cite{Bell2}: ``{\it It as only in the context of perfect correlation (or anticorrelation) that determinism could be inferred} [inferred, not assumed, Ed. note] {\it... Note well then that the following argument makes no mention whatever of determinism}''. Bell in fact presents  in this 1981 paper an alternative derivation of his inequalities, which dispenses with the EPR premise and with perfect anticorrelations altogether,  and relies only on Bell’s locality together with freedom of settings, to reach the same conclusion: no local theory can reproduce the observed quantum correlations.

The question then passes to experiment: do real systems obey or violate Bell inequalities? Pioneering tests in the 1970s~\cite{BTest1} and early 1980s~\cite{BTest2,BTest3}  observed violations consistent with Quantum Mechanics; later “loophole-free’’ experiments have confirmed this conclusion with increasingly stringent controls~\cite{BTest4,BTest5,BTest6}. This incredible experimental effort resulted in 2022 the Nobel Prize in Physics awarded to Aspect,  Clauser and Zeilinger for ``establishing the violation of Bell inequalities and pioneering quantum information science''.  

The upshot is clear: no local theory can account for the observed correlations. In Bell’s own terms, the world is \emph{nonlocal}—though in a way that still respects the no-signalling constraints of relativity, since these correlations cannot be used for superluminal communication. That tension remains conceptually striking: entangled, spacelike-separated systems display correlations that defy any local description, and this directly contradicts the hope, voiced in EPR, that a local completion of Quantum Mechanics might someday be found.

The practical consequences are profound. A measured Bell violation certifies the presence of nonclassical correlations \emph{from statistics alone}, enabling \emph{device-independent} (DI) guarantees in which security or correctness can be established without assuming a detailed or trusted model of the internal functioning of the devices involved. In quantum communication, this idea goes back to Ekert’s entanglement-based QKD~\cite{Ekert1991} and was put on a fully DI footing by works showing that observed nonlocality alone implies secrecy against general (even memory-bearing) adversaries~\cite{BHK2005,Acin2007,VaziraniVidick2014,ArnonFriedman2018}. These protocols are agnostic to vendor specifics: mismodelling of sources, detectors, or side channels cannot fake a Bell violation large enough to pass the security threshold so long as the spacetime and independence conditions enforced in the analysis are satisfied.  In networked scenarios, teleportation~\cite{Bennett1993} and entanglement swapping~\cite{Zukowski1993} serve as primitives for quantum repeaters~\cite{Briegel1998}. 
\emph{Teleportation} transfers an \emph{unknown} quantum state from Alice to Bob by \emph{consuming} a shared Bell pair and sending two classical bits that encode the outcome of a Bell-state measurement (BSM) performed by Alice; conditioned on those bits, Bob can reconstruct the input state, with no quantum signal traversing the channel during the transfer. 
\emph{Entanglement swapping} extends entanglement over distance: if nodes A--B and B--C each share an EPR pair, a BSM at the intermediate node B projects the remote systems at A and C into an entangled state---heralded by B’s classical outcome---even though A and C never interacted. 
\emph{Quantum repeaters} combine these basic building blocks to overcome the rapid degradation of quantum signals in direct transmission, allowing quantum correlations to be distributed reliably over long distances and thereby enabling tasks such as entanglement-based (and device-independent) QKD, clock and network synchronization, distributed sensing, and interconnects for modular quantum computing.

\section{Foundations for engineering}

Foundational results developed after Bell’s theorem proved equally powerful in practice and, crucially, capable of being combined and extended. Again, the central question is (non)locality and whether Quantum Theory admits a completion in terms of additional hidden variables. And again, Bell changed the game.

\emph{Contextuality} is the failure of a natural, classical-looking assumption: that the outcome assigned to an observable $A$ reflects a pre-existing property of the system \emph{alone} and is \emph{independent} of which other, compatible observables are measured alongside $A$ (the \emph{measurement context}). In a naïve—\emph{noncontextual}, in modern language—hidden-variable picture, one seeks a single value map $v$ that assigns predetermined outcomes to all observables once and for all. The assignment cannot be arbitrary: if a projective measurement is described by orthogonal projectors $\{P_i\}$ with $\sum_i P_i= I$, one demands $v(P_i)\in\{0,1\}$ and $\sum_i v(P_i)=1$, i.e., exactly one outcome “fires.” A series of no-go results~\cite{vonNeumann1932,JauchPiron1963,Gleason1957,KochenSpecker1967} makes this program untenable for Hilbert-space dimension $d>2$. For a time this was read as the end of any deeper (Einsteinian) description, despite the fact that a concrete completion—Bohmian mechanics—already existed.

Bell’s 1966 analysis clarified the landscape~\cite{Bell3}. He pinpointed which assumptions in early proofs were unwarranted. The price of a completion is that outcomes generally \emph{depend on the full measurement arrangement}—as in de~Broglie–Bohm theory, which is explicitly contextual (and nonlocal)~\cite{Bohm1952}. Ironically, it is  the orthodox quantum stance that saves hidden variables: measurement outcomes are not mere readouts of pre-existing properties but are defined by the \emph{interaction} of system and apparatus. Put differently, one cannot cleanly separate “the property of the system” from “the act of measurement”. The recorded result is a property of the joint system–context pair: {\it The result of an observation may reasonably depend not only on the state of the system (including hidden variables) but also on the complete disposition of the apparatus}~\cite{Bell3}.

Beyond foundations, this distinction has very practical consequences. A large and important class of quantum circuits, the \emph{stabilizer}  circuit---that is, a quantum circuit consisting solely of controlled-NOT (CNOT), Hadamard, and phase gates---can be simulated efficiently on a classical computer (Gottesman–Knill theorem). In other words, those dynamics admit a noncontextual  description and do not by themselves yield a computational advantage~\cite{Gottesman1997,AaronsonGottesman2004}. To go beyond that regime one must inject new ingredients, typically in the form of specially prepared \emph{magic states}. These states fail to admit any noncontextual  model; they exhibit interference patterns that classical samplers cannot track efficiently. This is why contextuality is now viewed as a \emph{fuel} for quantum computation~\cite{Howard2014,Raussendorf2013}. In short, Bell’s 1966 lesson—hidden variables are not excluded \emph{per se}, only \emph{noncontextual} ones are—maps cleanly onto engineering: to outrun classically simulable stabilizer dynamics, a scalable device must supply a steady throughput of  magic states and preserve enough coherence for them to do useful work.

The \emph{no-cloning} theorem~\cite{WoottersZurek1982,Dieks1982} and its mixed-state extension, \emph{no-broadcasting}~\cite{Barnum1996}  forbid  making perfect copies of unknown quantum states. From a foundational standpoint, no-cloning is crucial because it blocks superluminal signalling: proposals such as Herbert’s FLASH device~\cite{Herbert1982} attempted to exploit entanglement plus hypothetical cloning to transmit information faster than light, but no-cloning closes that route. No-cloning and no-broadcasting then migrated from conceptual curiosities to architectural constraints: they forbid naive signal amplification in long-distance links, thereby motivating repeater architectures and error-corrected memories; in cryptography, they underwrite the intuition that adversaries cannot make perfect copies of unknown quantum states (a pillar behind prepare-and-measure and entanglement-based QKD). 

Finally, the logic of \emph{quantum error correction} reconciles fragility of quantum states  with scalability by encoding information nonlocally so that local noise induces detectable signatures: stabilizer and CSS codes~\cite{Shor1995,Steane1996,Gottesman1997} and topological/surface codes~\cite{Dennis2002,Fowler2012} make faults \emph{identifiable} and \emph{correctable} while preserving global coherence, and threshold theorems~\cite{AharonovBenOr1997,Aliferis2006} guarantee reliability below a constant physical error rate.

\section{From many to one, and back to many}

Schr\"odinger, not only Einstein, was deeply uneasy about the status of Quantum Mechanics as a theory of nature. His famous “cat’’ paradox~\cite{Schrodinger1935} was meant precisely to dramatize the oddity of applying the linear superposition principle to macroscopic objects: if the theory is taken at face value for individual systems, it seems to predict superpositions of mutually exclusive macroscopic states, which clashes with our experience. The same unease surfaces in his reflections on experimental practice: \emph{“We} never {\it experiment with just} one {\it electron or atom or (small) molecule. In thought-experiments we sometimes assume that we do; this invariably entails ridiculous consequences”}~\cite{Schrodinger1952}. Read charitably and in context, these statements capture a widespread 1930s–1950s view: Quantum Mechanics was exquisitely successful as a calculus for \emph{ensembles}—be it ensembles of atoms in condensed-matter systems or of photons in laser physics—while genuine single-particle preparation and control seemed unattainable. 

The ensuing history has, of course, rewritten that expectation. Starting from ensembles, advances in laser cooling and trapping put individual atoms at our fingertips: the 1990s toolbox (laser cooling, sub-Doppler techniques, optical molasses and magneto-optical traps) is exemplified by the first MOT~\cite{Raab1987} and by the observation and theory of (sub-Doppler) cooling~\cite{Lett1988,Dalibard1989}. Within a few years, ensembles were driven across a new frontier with the first Bose–Einstein condensates in dilute gases (rubidium at JILA; sodium at MIT), demonstrating macroscopic matter-wave coherence~\cite{Anderson1995,Davis1995}. In parallel on the photonic side, quantum optics matured from ensemble descriptions of fields to exquisitely controlled nonclassical light: Glauber’s 1963 papers provided the Quantum Theory of optical coherence and photon statistics~\cite{Glauber1963a,Glauber1963b}, while the advent of  \emph{frequency combs} linked optical and microwave frequencies with phase-coherent precision, enabling modern optical clocks and precision metrology~\cite{Diddams2000PRL,Holzwarth2000PRL}. 

The conceptual leap from “many’’ to “one’’ became reality with direct manipulation of \emph{individual} quantum systems: cavity-QED experiments “saw’’ single photons without destroying them and tracked their quantum jumps~\cite{Nogues1999,Gleyzes2007}, while trapped-ion platforms observed quantum jumps in a single ion and achieved ground-state cooling, enabling coherent single-quanta control~\cite{Nagourney1986,Sauter1986,Diedrich1989,Monroe1995}. Foundationally, this was a turning point: the single-system regime that Schr\"odinger had regarded as practically out of reach became a working laboratory, where superposition, entanglement, and measurement back-action could be prepared, witnessed, and recycled on demand. 

With those tools in hand, the field effectively “closed the loop’’: having learned to isolate and manipulate single atoms and photons, researchers began to assemble controlled \emph{few-} and \emph{many-body} structures from the bottom up, with the explicit aim of realizing programmable quantum simulators and scalable quantum information processors. Optical lattices loaded with ultracold atoms enabled faithful realizations of Hubbard-model physics with tunable interactions, disorder, and dimensionality; quantum-gas microscopes subsequently pushed these systems to single-site resolution and control, turning textbook lattice Hamiltonians into directly imageable and programmable matter-wave circuits~\cite{BlochRMP2008,Bakr2009,Sherson2010}. In parallel, Rydberg-atom arrays assembled by optical tweezers provided defect-free registers with strong, controllable interactions and native entangling gates, while trapped-ion chains and, more recently, high-cooperativity cavity and waveguide QED platforms offered complementary routes to scalable entanglement generation, coherent control, and mid-circuit measurement~\cite{Saffman2010RMP,Browaeys2020,BlattWineland2008,ReisererRempe2015,Lodahl2015}. On the photonic front, integrated waveguides and deterministic emitters (semiconductor quantum dots~\cite{Senellart2017}; color centers in solids~\cite{Aharonovich2016}) have transformed “single-photon’’ generation from a conceptual milestone into an engineering specification, enabling boson-sampling tests of computational hardness and the implementation of measurement-based quantum computation with cluster states~\cite{Brod2019,RaussendorfBriegel2001}.

Across these developments, two broad architectural paradigms have emerged for converting single-particle control into scalable quantum information processing and controlled many-body dynamics. Atomic platforms encode qubits in long-lived internal states and exploit interactions that are either mediated collectively, as in trapped-ion processors where entangling gates arise from shared motional modes, or engineered locally, as in neutral-atom systems combining the massive parallelism of quantum-gas microscopes with the programmability of optical tweezers. Photonic platforms follow a complementary strategy: rather than relying on direct interactions, they achieve universality through linear optics, high-quality single-photon sources, and photon counting, where measurement and feedforward promote passive interferometers into fully programmable quantum processors, equivalently described in the cluster-state, measurement-based paradigm.

Conceptually, these machines instantiate Feynman’s  program~\cite{Feynman1982}: controllable quantum hardware configured to emulate quantum dynamics beyond classical reach—either by \emph{programming gates} (digital) or by \emph{tuning interactions} (analog)—with foundations (entanglement, measurement back-action, and nonclassical interference) now functioning as design constraints and resources.

This progression—from ensembles to singles and back to engineered many-body systems—is not merely technical; it has reshaped how foundations and technology inform one another. Technology is now pushing toward the creation and control of ever larger quantum states. Schr\"odinger’s cat remains a compelling parable, yet in the laboratory we routinely prepare complex superpositions (involving many particles or many photons), monitor their decoherence channels in real time, stabilize them with measurement-based feedback and error correction, and quantify how “nonclassical’’ they are through operational witnesses tailored to specific tasks.

\section{The math of quantum technologies}

Quantum technologies ultimately draw their power—and encounter their fundamental limitations—from the mathematical structure of Quantum Theory itself. While many of the phenomena discussed so far can be described operationally or experimentally, a precise understanding of what can and cannot be achieved requires engaging directly with the formal language of states, transformations, and measurements. We now briefly introduce the minimal mathematical framework needed to articulate these constraints, showing how abstract structural features of Quantum Theory translate into concrete consequences for information processing, control, and scalability.

A compact, operational formulation of Quantum Theory emerges in the early '30s from the work of Dirac and von~Neumann. Dirac’s \emph{Principles}~\cite{Dirac1930} introduced the bra–ket calculus and spectral expansions in physics language, identifying observables with Hermitian operators on a complex Hilbert space $\mathcal H$ and amplitudes with inner products. Von~Neumann then supplied the functional-analytic backbone~\cite{vonNeumann1932}: (i) states are positive trace-one operators $\rho$; (ii) sharp measurements are projection-valued measures (PVMs) obtained from the spectral theorem for self-adjoint operators; (iii) expectation values follow Born’s rule $\langle A\rangle=\mathrm{Tr}(\rho A)$; and (iv) closed dynamics are unitary; in this ideal limit, pure states obey Schr\"odinger’s equation and mixed states the von~Neumann equation $\dot\rho=-\,\mathrm i[H,\rho]$.

Ideal \emph{projective} measurements of an observable $A=\sum_a a\,\Pi_a$ produce outcomes with probabilities $\Pr(a)=\mathrm{Tr}(\rho\,\Pi_a)$ and update the state via the L\"uders rule
$ \rho\ \mapsto\ \rho_a=\Pi_a \rho \Pi_a/\mathrm{Tr}(\rho \Pi_a)$~\cite{Lueders1951}. When the readout couples to an integral of motion (or a commuting observable), projective measurements can be \emph{quantum non-demolition} (QND)~\cite{Caves1980,BraginskyKhalili1992}. Real detectors, however, are seldom ideal projectors. The general statistical structure is a Positive Operator-Valued Measure (POVM): a set of effects $\{E_x\}$ with $E_x\!\ge0$ and $\sum_x E_x=\openone$, giving $\Pr(x)=\mathrm{Tr}(\rho E_x)$~\cite{DaviesLewis1970,Holevo1973}. Naimark’s dilation theorem shows any POVM arises as a projective measurement on a larger space~\cite{Naimark1940}.

This is all true for isolated quantum systems. However, realistic devices interact with their surroundings. Stinespring’s dilation (and Kraus/Choi representations) characterizes the most general physical evolution as a completely positive, trace-preserving (CPTP) map
$
\Phi(\rho)=\sum_k K_k \rho K_k^\dagger$, with  $\sum_k K_k^\dagger K_k=\openone
$~\cite{Stinespring1955,Kraus1971,Choi1975}. When the map is the result of a  (Markovian) continuous time evolution $\Lambda_t$, the family $\{\Lambda_t\}_{t\ge0}$ forms a quantum dynamical semigroup with generator in \emph{GKSL} (Gorini-Kossakowski-Sudarshan/Linblad) form,
\[
\dot\rho=\mathcal L(\rho)=-\,\mathrm i[H,\rho]+\sum_j \gamma_j\!\left(L_j\rho L_j^\dagger-\frac{1}{2}\{L_j^\dagger L_j,\rho\}\right),
\]
the unique linear structure compatible with complete positivity, trace preservation, and semigroup composition~\cite{Lindblad1976,GKS1976,Davies1974,Davies1976,Spohn1980}. This form makes explicit that the reduced dynamics of an open quantum system is fully characterized by a competition between coherent unitary evolution and irreversible dissipative processes, the latter encoding both decoherence and noise through a structure that is fixed by complete positivity and trace preservation rather than by microscopic details.

This open-systems calculus~\cite{BreuerPetruccione2002} has the great merit of explaining and quantifying {\it decoherence}: why quantum properties are  progressively lost when systems interact with the surrounding environment.  Quantum Brownian motion models, pioneered by Caldeira and Leggett~\cite{CaldeiraLeggett1983,LeggettRMP1987} and later extended to include collisional decoherence~\cite{JoosZeh1985} and exact master-equation formulations~\cite{HuPazZhang1992}, are now routinely employed to assess the impact of external and internal noise in experiments.

Between ideal projective and general POVM readouts lies the regime of \emph{weak measurements}~\cite{AharonovAlbertVaidman1988,DuckSudarshan1989,Dressel2014}, which enable gentle monitoring of dynamics, direct retrieval of phases and amplitudes, and—in certain technical-noise regimes—weak-value amplification of detector signals~\cite{Lundeen2011,HostenKwiat2008,Dixon2009,Jordan2014}. Taking the continuous-time limit yields \emph{continuous quantum measurement}. Belavkin’s treatment conditions the system state on a measurement record (photon counts, homodyne currents), producing stochastic master equations (SMEs) driven by Poisson or Wiener processes; Hudson–Parthasarathy quantum stochastic calculus derive the same SMEs from field couplings~\cite{Belavkin1992,HudsonParthasarathy1984,GardinerCollett1985,JacobsSteck2006}. “Quantum trajectories’’ then unravel GKSL dynamics into single-shot state paths whose ensemble average reproduces the unconditional evolution~\cite{DalibardCastinMolmer1992,Carmichael1993,WisemanMilburn1993}. 

These ingenious mathematical structures have redefined laboratory capability. In {\it circuit QED}, dispersive QND readout produces a homodyne current carrying information about a qubit Bloch component; the SME quantifies information gain vs.\ backaction, enabling real-time feedback to stabilize Rabi oscillations, and even \emph{catch and reverse} an impending quantum jump~\cite{Vijay2012,Weber2014,Murch2013,Hatridge2013,Minev2019}. \emph{In cavity and levitated optomechanics,} continuous position readout via homodyne detection implements the same SME toolkit: measurement-based feedback and optimal control cool mechanical motion to (or near) the quantum ground state and stabilize trajectories in real time, turning the standard-quantum-limit trade-off between imprecision and backaction into an engineered resource~\cite{Aspelmeyer2014RMP,Caves1980,Delic2020Science,Magrini2021Nature,Tebbenjohanns2021Nature}. Backaction-evading, single-quadrature measurements and QND strategies for mechanical observables further enable sub-SQL (Standard Quantum Limit) force sensing and state preparation, with squeezed/filtered probes and high-cooperativity cavities suppressing added noise while preserving coherence for subsequent tasks.

Thus the same mathematics—POVMs and dilations, CPTP channels and semigroups, weak and continuous measurement SMEs—both grounds foundational debates and provides the design language for devices whose performance can be predicted, certified, and improved.

\section{Technology probing foundations}

Two near-term scientific frontiers show how technology now pushes back on foundations. 
\emph{First: the assessment of quantumness of gravity.}  Gravity is not just another force on a fixed stage—it \emph{is} the stage: the geometry of space and time. Turning that stage itself into a quantum object is conceptually challenging, and the usual techniques that tame infinities in other quantum field theories fail for gravity. A halfway, semiclassical picture—classical spacetime guided by quantum matter—sidesteps the deepest question: what happens when matter is in a superposition of masses or positions? These tensions motivate operational tests that can reveal whether gravity is genuinely quantum, without committing to any single theory.

Away from the Planck scale, where quantum and gravitational phenomena are expected to be equally important but remain experimentally inaccessible, a clean low-energy program asks whether gravity can act as a \emph{quantum mediator}. The core idea is to prepare two mesoscopic masses in spatially separated superpositions and let them interact \emph{only} via gravity; if entanglement appears, then the mediator must be quantum~\cite{Bose2017,MarlettoVedral2017}. The challenge is fundamental: gravitational interactions become stronger for larger masses and separations, while quantum control is most effective for smaller, colder, and more isolated objects. Reaching the “sweet spot’’ demands aggressive gains in isolation, cryogenics, ground- and vibration-noise suppression, charge control and patch-potential mitigation, and quantum-limited readout. Reaching the “sweet spot’’ demands aggressive gains in isolation, cryogenics, ground- and vibration-noise suppression, charge control and patch-potential mitigation, and quantum-limited readout. Steady progress in cavity/levitated optomechanics~\cite{Aspelmeyer2014RMP} and precision control has moved this from gedanken to campaign: ground-state cooling and quantum control of mechanical motion are now routine goals for nanoparticles and membranes~\cite{Delic2020Science,Magrini2021Nature,Tebbenjohanns2021Nature}.  A positive observation would be the first tabletop evidence for gravity’s quantum character; a sustained null result over a carefully delimited parameter window would prune semiclassical alternatives~\cite{Kafri2014,Tilloy2018,Oppenheim2023} and redirect designs toward architectures where gravitation is operationally classical at relevant scales. 

\emph{Second: objective-collapse dynamics.}  In standard Quantum Mechanics, closed systems evolve linearly and unitarily by Schr\"odinger dynamics, yet measurements yield single, definite outcomes with Born-rule probabilities. When a system $S$ interacts with an apparatus $A$, the joint evolution produces an $S$–$A$ entangled superposition; the projection postulate (L\"uders/von~Neumann update) then stipulates a nonunitary, stochastic ``collapse’’ yielding a definite outcome. The collapse however is tied to an imprecise notion of measurement, raising the question when a configuration can be called a measurement---causing the collapse of the wave function---and when it can not~\cite{BellSp}. These tensions define the measurement problem in Quantum Mechanics and have led to several proposed resolutions, including objective-collapse models (GRW/CSL), hidden-variable theories (de Broglie–Bohm), and many-worlds approaches. We focus below on collapse models, which offer experimentally testable departures from Schrödinger evolution.

Spontaneous-collapse models (GRW/CSL and variants) posit a universal, weak, mass-weighted noise that localizes the wave function in position~\cite{GRW1986,Pearle1989,GPR1990,Bassi2013RMP}; they predict small, testable departures from linear Schr\"odinger evolution that grow with system size. Two complementary strategies are closing in. Interferometric tests~\cite{ArndtHornberger2014} push matter-wave coherence to higher masses and larger path separations, hunting for loss of visibility that \emph{exceeds} environmental decoherence—progress spanning molecular interferometry to proposals for macroscopic superpositions with opto- and magneto-mechanical objects~\cite{Hornberger2012RMP,Gerlich2011,Fein2019,Bateman2014}. Noninterferometric tests~\cite{Carlesso2022NatPhys} search for the universal momentum diffusion and spontaneous radiation implied by CSL: ultralow-noise cantilevers, levitated particles, and cryogenic resonators set bounds from excess heating and force noise~\cite{Vinante2016,Vinante2017,Pontin2020}, while X-ray emission constraints and bulk-heating analyses bound the collapse rate in complementary parameter regions~\cite{Fu1997,Arnquist2022}. 

The moral in both frontiers is the same: the very techniques that power quantum technologies—ultrahigh vacuum, cryogenics, laser/charge control, ground-state cooling, quantum-limited detection, and meticulous system identification—turn deep foundational questions into empirical targets with credible error bars. Culture matters too: where the twentieth century sometimes caricatured foundations as metaphysics, the twenty-first treats them as part of physics, susceptible to theoretical and experimental investigations.

\section*{Conclusions}
Foundations are not a preface to technology; they are its operating system. The same features of Quantum Theory that long appeared paradoxical—superposition, entanglement, contextuality—are precisely those that enable technological advantage when they are stabilized, controlled, and certified. The counterintuitive features of Quantum Mechanics that capture public imagination are the very same features that enable new forms of information processing and technological advantage. Bell’s migration from philosophy to inequalities taught us how to turn conceptual questions into experimental benchmarks; the maturation of quantum control now allows us to run that logic in reverse, using advanced platforms to probe first principles themselves—whether linear Quantum Theory is exact at all relevant scales, whether gravity can mediate entanglement, and where the true quantum limits lie. Seen in this light, the International Year of Quantum (2025) is not only a celebration, but an invitation to let our most sophisticated devices answer our deepest questions. The dividends are twofold: for science, experiments that move long-standing debates from the speculative to the empirical; for technology, the continued drive toward more powerful devices and protocols. Ultimately, the future of Quantum Theory will be limited only by how boldly we allow foundational insight to shape it.
\\


\noindent {\bf Acknowledgements.} A.B. acknowledges financial support from the EIC Pathfinder project QuCoM (GA no. 101046973), the PNRR PE National Quantum Science and Technology Institute (GA no. PE0000023), the University of Trieste and
INFN. He thanks Sandro Donadi and Matteo Marinelli  for comments of earlier drafts of the manuscript.


\begin{thebibliography}{999}

\bibitem{BohrEinstein}
Jammer M 1974 \emph{The Philosophy of Quantum Mechanics: The Interpretations of Quantum Mechanics in Historical Perspective} (New York: Wiley)

\bibitem{Howard}
Howard D 1990 ‘Nicht Sein Kann Was Nicht Sein Darf,’ or the prehistory of EPR, 1909–1935: Einstein’s early worries about the quantum mechanics of composite systems in \emph{Sixty-Two Years of Uncertainty: Historical, Philosophical, and Physical Inquiries into the Foundations of Quantum Mechanics} ed A I Miller (NATO ASI Series vol 226) (New York: Plenum) pp 61–111

\bibitem{solvay}
Bacciagaluppi G and Valentini A 2009 \emph{Quantum Theory at the Crossroads}, section ``General discussion of the new ideas presented'' (Cambridge: Cambridge Univ. Press)

\bibitem{epr}
Einstein A, Podolsky B and Rosen N 1935 Can quantum-mechanical description of physical reality be considered complete? \emph{Phys. Rev.} \textbf{47} 777–780

\bibitem{Bohr}
Bohr N 1935 Can quantum-mechanical description of physical reality be considered complete? \emph{Phys. Rev.} \textbf{48} 696–702

\bibitem{Pauli}
Einstein A and Born M 1971 \emph{The Born–Einstein Letters: Correspondence between Albert Einstein and Max and Hedwig Born from 1916 to 1955} (New York: Walker; London: Macmillan) [includes Pauli’s 31 Mar 1954 letter to Born]

\bibitem{Bohm}
Bohm D 1951 \emph{Quantum Theory} (Englewood Cliffs, NJ: Prentice-Hall); reprinted 1989 (Mineola, NY: Dover)

\bibitem{Bell1}
Bell J S 1964 On the Einstein Podolsky Rosen paradox \emph{Physics Physique Fizika} \textbf{1} 195–200. Reprinted in~\cite{BellSp}.

\bibitem{BellSp}
Bell J S 1987 \emph{Speakable and Unspeakable in Quantum Mechanics} (Cambridge: Cambridge Univ. Press) 2nd edn 2004

\bibitem{deBroglie1924}
de Broglie L 1925 Recherches sur la théorie des quanta \emph{Ann. Phys.}  \textbf{10} 22–128

\bibitem{Bohm1952}
Bohm D 1952 A suggested interpretation of the quantum theory in terms of “hidden’’ variables I and II \emph{Phys. Rev.} \textbf{85} 166–179; 180–193

\bibitem{DurrTeufel2009}
D\"urr D and Teufel S 2009 \emph{Bohmian Mechanics: The Physics and Mathematics of Quantum Theory} (Berlin: Springer)

\bibitem{Bell2}
Bell J S 1981 Bertlmann’s socks and the nature of reality \emph{J. Phys. Colloq.} \textbf{42} C2-41. Reprinted in~\cite{BellSp}.

\bibitem{BTest1}
Freedman S J and Clauser J F 1972 Experimental test of local hidden-variable theories \emph{Phys. Rev. Lett.} \textbf{28} 938–941

\bibitem{BTest2}
Aspect A, Grangier P and Roger G 1981 Experimental tests of realistic local theories via Bell’s theorem \emph{Phys. Rev. Lett.} \textbf{47} 460–463

\bibitem{BTest3}
Aspect A, Dalibard J and Roger G 1982 Experimental test of Bell’s inequalities using time-varying analyzers \emph{Phys. Rev. Lett.} \textbf{49} 1804–1807

\bibitem{BTest4}
Hensen B \emph{et al.} 2015 Loophole-free Bell inequality violation using electron spins separated by 1.3 kilometres \emph{Nature} \textbf{526} 682–686

\bibitem{BTest5}
Giustina M \emph{et al.} 2015 Significant-loophole-free test of Bell’s theorem with entangled photons \emph{Phys. Rev. Lett.} \textbf{115} 250401

\bibitem{BTest6}
Shalm L K \emph{et al.} 2015 Strong loophole-free test of local realism \emph{Phys. Rev. Lett.} \textbf{115} 250402

\bibitem{Ekert1991}
Ekert A K 1991 Quantum cryptography based on Bell’s theorem \emph{Phys. Rev. Lett.} \textbf{67} 661–663

\bibitem{BHK2005}
Barrett J, Hardy L and Kent A 2005 No signaling and quantum key distribution \emph{Phys. Rev. Lett.} \textbf{95} 010503

\bibitem{Acin2007}
Acín A, Brunner N, Gisin N, Massar S, Pironio S and Scarani V 2007 Device-independent security of quantum cryptography against collective attacks \emph{Phys. Rev. Lett.} \textbf{98} 230501

\bibitem{VaziraniVidick2014}
Vazirani U and Vidick T 2014 Fully device-independent quantum key distribution \emph{Phys. Rev. Lett.} \textbf{113} 140501

\bibitem{ArnonFriedman2018}
Arnon-Friedman R, Dupuis F, Fawzi O, Renner R and Vidick T 2018 Simple and tight device-independent security proofs via entropy accumulation \emph{Nat. Commun.} \textbf{9} 459

\bibitem{Bennett1993}
Bennett C H, Brassard G, Crépeau C, Jozsa R, Peres A and Wootters W K 1993 Teleporting an unknown quantum state via dual classical and Einstein–Podolsky–Rosen channels \emph{Phys. Rev. Lett.} \textbf{70} 1895–1899

\bibitem{Zukowski1993}
\.{Z}ukowski M, Zeilinger A, Horne M A and Ekert A K 1993 ‘Event-ready-detectors’ Bell experiment via entanglement swapping \emph{Phys. Rev. Lett.} \textbf{71} 4287–4290

\bibitem{Briegel1998}
Briegel H-J, Dür W, Cirac J I and Zoller P 1998 Quantum repeaters: the role of imperfect local operations in quantum communication \emph{Phys. Rev. Lett.} \textbf{81} 5932–5935

\bibitem{vonNeumann1932}
von Neumann J 1932 \emph{Mathematische Grundlagen der Quantenmechanik} (Berlin: Springer); English transl. Beyer R T 1955 \emph{Mathematical Foundations of Quantum Mechanics} (Princeton, NJ: Princeton Univ. Press)

\bibitem{JauchPiron1963}
Jauch J M and Piron C 1963 Can hidden variables be excluded in quantum mechanics? \emph{Helv. Phys. Acta} \textbf{36} 827–837

\bibitem{Gleason1957}
Gleason A M 1957 Measures on the closed subspaces of a Hilbert space \emph{J. Math. Mech.} \textbf{6} 885–893

\bibitem{KochenSpecker1967}
Kochen S and Specker E P 1967 The problem of hidden variables in quantum mechanics \emph{J. Math. Mech.} \textbf{17} 59–87

\bibitem{Bell3}
Bell J S 1966 On the problem of hidden variables in quantum mechanics \emph{Rev. Mod. Phys.} \textbf{38} 447–452. Reprinted in~\cite{BellSp}.

\bibitem{Gottesman1997}
Gottesman D 1997 \emph{Stabilizer Codes and Quantum Error Correction} (PhD thesis, California Institute of Technology). ArXiv:quant-ph/9705052

\bibitem{AaronsonGottesman2004}
Aaronson S and Gottesman D 2004 Improved simulation of stabilizer circuits \emph{Phys. Rev. A} \textbf{70} 052328

\bibitem{Howard2014}
Howard M, Wallman J J, Veitch V and Emerson J 2014 Contextuality supplies the ‘magic’ for quantum computation \emph{Nature} \textbf{510} 351–355

\bibitem{Raussendorf2013}
Raussendorf R 2013 Contextuality in measurement-based quantum computation \emph{Phys. Rev. A} \textbf{88} 022322

\bibitem{WoottersZurek1982}
Wootters W K and Zurek W H 1982 A single quantum cannot be cloned \emph{Nature} \textbf{299} 802–803

\bibitem{Dieks1982}
Dieks D 1982 Communication by EPR devices \emph{Phys. Lett. A} \textbf{92} 271–272

\bibitem{Barnum1996}
Barnum H, Caves C M, Fuchs C A, Jozsa R and Schumacher B 1996 Noncommuting mixed states cannot be broadcast \emph{Phys. Rev. Lett.} \textbf{76} 2818–2821

\bibitem{Herbert1982}
Herbert N 1982 FLASH—A superluminal communicator based upon a new kind of quantum measurement \emph{Found. Phys.} \textbf{12} 1171–1179

\bibitem{Shor1995}
Shor P W 1995 Scheme for reducing decoherence in quantum computer memory \emph{Phys. Rev. A} \textbf{52} R2493–R2496

\bibitem{Steane1996}
Steane A M 1996 Error correcting codes in quantum theory \emph{Phys. Rev. Lett.} \textbf{77} 793–797

\bibitem{Dennis2002}
Dennis E, Kitaev A, Landahl A and Preskill J 2002 Topological quantum memory \emph{J. Math. Phys.} \textbf{43} 4452–4505

\bibitem{Fowler2012}
Fowler A G, Mariantoni M, Martinis J M and Cleland A N 2012 Surface codes: Towards practical large-scale quantum computation \emph{Phys. Rev. A} \textbf{86} 032324

\bibitem{AharonovBenOr1997}
Aharonov D and Ben-Or M 1997 Fault-tolerant quantum computation with constant error in \emph{Proc. 29th Annual ACM Symposium on Theory of Computing (STOC ’97)} (New York: ACM) pp 176–188

\bibitem{Aliferis2006}
Aliferis P, Gottesman D and Preskill J 2006 Quantum accuracy threshold for concatenated distance-3 codes \emph{Quantum Inf. Comput.} \textbf{6} 97–165

\bibitem{Schrodinger1935}
Schrödinger E 1935 Die gegenwärtige Situation in der Quantenmechanik \emph{Die Naturwissenschaften} \textbf{23} 807–812; 823–828; 844–849 [Engl. transl. 1980 \emph{Proc. Am. Phil. Soc.} \textbf{124} 323–338 The present situation in quantum mechanics]

\bibitem{Schrodinger1952}
Schrödinger E 1952 Are there quantum jumps? Part II \emph{Br. J. Philos. Sci.} \textbf{3} 233–242

\bibitem{Raab1987}
Raab E L, Prentiss M, Cable A, Chu S and Pritchard D E 1987 Trapping of neutral sodium atoms with radiation pressure \emph{Phys. Rev. Lett.} \textbf{59} 2631–2634

\bibitem{Lett1988}
Lett P D, Watts R N, Westbrook C I, Phillips W D, Gould P L and Metcalf H J 1988 Observation of atoms laser cooled below the Doppler limit \emph{Phys. Rev. Lett.} \textbf{61} 169–172

\bibitem{Dalibard1989}
Dalibard J and Cohen-Tannoudji C 1989 Laser cooling below the Doppler limit by polarization gradients: simple theoretical models \emph{J. Opt. Soc. Am. B} \textbf{6} 2023–2045

\bibitem{Anderson1995}
Anderson M H, Ensher J R, Matthews M R, Wieman C E and Cornell E A 1995 Observation of Bose–Einstein condensation in a dilute atomic vapor \emph{Science} \textbf{269} 198–201

\bibitem{Davis1995}
Davis K B, Mewes M-O, Andrews M R, van Druten N J, Durfee D S, Kurn D M and Ketterle W 1995 Bose–Einstein condensation in a gas of sodium atoms \emph{Phys. Rev. Lett.} \textbf{75} 3969–3973

\bibitem{Glauber1963a}
Glauber R J 1963 The quantum theory of optical coherence \emph{Phys. Rev.} \textbf{130} 2529–2539

\bibitem{Glauber1963b}
Glauber R J 1963 Coherent and incoherent states of the radiation field \emph{Phys. Rev.} \textbf{131} 2766–2788

\bibitem{Diddams2000PRL}
Diddams S A, Jones D J, Ye J, Cundiff S T, Hall J L, Ranka J K, Windeler R S, Holzwarth R, Udem T and Hänsch T W 2000 Direct link between microwave and optical frequencies with a 300 THz femtosecond laser comb \emph{Phys. Rev. Lett.} \textbf{84} 5102–5105

\bibitem{Holzwarth2000PRL}
Holzwarth R, Udem T, Hänsch T W, Knight J C, Wadsworth W J and Russell P S J 2000 Optical frequency synthesizer for precision spectroscopy \emph{Phys. Rev. Lett.} \textbf{85} 2264–2267

\bibitem{Nogues1999}
Nogues G, Rauschenbeutel A, Osnaghi S, Brune M, Raimond J M and Haroche S 1999 Seeing a single photon without destroying it \emph{Nature} \textbf{400} 239–242

\bibitem{Gleyzes2007}
Gleyzes S, Kuhr S, Guerlin C, Bernu J, Deléglise S, Hoff U B, Brune M, Raimond J-M and Haroche S 2007 Quantum jumps of light: recording the birth and death of a photon in a cavity \emph{Nature} \textbf{446} 297–300

\bibitem{Nagourney1986}
Nagourney W, Sandberg J and Dehmelt H 1986 Shelved optical electron amplifier: observation of quantum jumps \emph{Phys. Rev. Lett.} \textbf{56} 2797–2799

\bibitem{Sauter1986}
Sauter T, Neuhauser W, Blatt R and Toschek P E 1986 Observation of quantum jumps \emph{Phys. Rev. Lett.} \textbf{57} 1696–1699

\bibitem{Diedrich1989}
Diedrich F, Bergquist J C, Itano W M and Wineland D J 1989 Laser cooling to the zero-point energy of motion \emph{Phys. Rev. Lett.} \textbf{62} 403–406

\bibitem{Monroe1995}
Monroe C, Meekhof D M, King B E, Itano W M and Wineland D J 1995 Demonstration of a fundamental quantum logic gate \emph{Phys. Rev. Lett.} \textbf{75} 4714–4717

\bibitem{BlochRMP2008}
Bloch I, Dalibard J and Zwerger W 2008 Many-body physics with ultracold gases \emph{Rev. Mod. Phys.} \textbf{80} 885–964

\bibitem{Bakr2009}
Bakr W S, Gillen J I, Peng A, Fölling S and Greiner M 2009 A quantum gas microscope for detecting single atoms in a Hubbard-regime optical lattice \emph{Nature} \textbf{462} 74–77

\bibitem{Sherson2010}
Sherson J F, Weitenberg C, Endres M, Cheneau M, Bloch I and Kuhr S 2010 Single-atom-resolved fluorescence imaging of an atomic Mott insulator \emph{Nature} \textbf{467} 68–72

\bibitem{Saffman2010RMP}
Saffman M, Walker T G and Mølmer K 2010 Quantum information with Rydberg atoms \emph{Rev. Mod. Phys.} \textbf{82} 2313–2363

\bibitem{Browaeys2020}
Browaeys A and Lahaye T 2020 Many-body physics with individually controlled Rydberg atoms \emph{Nat. Phys.} \textbf{16} 132–142

\bibitem{BlattWineland2008}
Blatt R and Wineland D J 2008 Entangled states of trapped atomic ions \emph{Nature} \textbf{453} 1008–1015

\bibitem{ReisererRempe2015}
Reiserer A and Rempe G 2015 Cavity-based quantum networks with single atoms and optical photons \emph{Rev. Mod. Phys.} \textbf{87} 1379–1418

\bibitem{Lodahl2015}
Lodahl P, Mahmoodian S and Stobbe S 2015 Interfacing single photons and single quantum dots with photonic nanostructures \emph{Rev. Mod. Phys.} \textbf{87} 347–400

\bibitem{Senellart2017}
Senellart P, Solomon G and White A 2017 High-performance semiconductor quantum-dot single-photon sources \emph{Nat. Nanotechnol.} \textbf{12} 1026–1039

\bibitem{Aharonovich2016}
Aharonovich I, Englund D and Toth M 2016 Solid-state single-photon emitters \emph{Nat. Photonics} \textbf{10} 631–641

\bibitem{Brod2019}
Brod D J, Galvão E F, Crespi A, Osellame R, Spagnolo N and Sciarrino F 2019 Photonic implementation of boson sampling: a review \emph{Adv. Photonics} \textbf{1} 034001

\bibitem{RaussendorfBriegel2001}
Raussendorf R and Briegel H J 2001 A one-way quantum computer \emph{Phys. Rev. Lett.} \textbf{86} 5188–5191

\bibitem{Feynman1982}
Feynman R P 1982 Simulating physics with computers \emph{Int. J. Theor. Phys.} \textbf{21} 467–488

\bibitem{Lloyd1996}
Lloyd S 1996 Universal quantum simulators \emph{Science} \textbf{273} 1073–1078

\bibitem{Georgescu2014}
Georgescu I M, Ashhab S and Nori F 2014 Quantum simulation \emph{Rev. Mod. Phys.} \textbf{86} 153–185

\bibitem{Dirac1930}
Dirac P A M 1930 \emph{The Principles of Quantum Mechanics} (Oxford: Clarendon Press) [4th edn 1958]

\bibitem{Lueders1951}
Lüders G 1950 Über die Zustandsänderung durch den Meßprozeß \emph{Ann. Phys.} \textbf{443} 322–328

\bibitem{Caves1980}
Caves C M, Thorne K S, Drever R W P, Sandberg V D and Zimmermann M 1980 On the measurement of a weak classical force coupled to a quantum-mechanical oscillator. I. Issues of principle \emph{Rev. Mod. Phys.} \textbf{52} 341–392

\bibitem{BraginskyKhalili1992}
Braginsky V B, Khalili F Y and Thorne K S 1992 \emph{Quantum Measurement} (Cambridge: Cambridge Univ. Press)

\bibitem{DaviesLewis1970}
Davies E B and Lewis J T 1970 An operational approach to quantum probability \emph{Commun. Math. Phys.} \textbf{17} 239–260

\bibitem{Holevo1973}
Holevo A S 1973 Statistical decision theory for quantum systems \emph{J. Multivar. Anal.} \textbf{3} 337–394

\bibitem{Naimark1940}
Naimark M A 1940 Spectral functions of a symmetric operator \emph{Izv. Akad. Nauk SSSR Ser. Mat.} \textbf{4} 277–318 [Engl. transl.: \emph{Amer. Math. Soc. Transl.} (2) \textbf{16} 103–193 (1960)]

\bibitem{Stinespring1955}
Stinespring W F 1955 Positive functions on $C^\ast$-algebras \emph{Proc. Amer. Math. Soc.} \textbf{6} 211–216

\bibitem{Kraus1971}
Kraus K 1971 General state changes in quantum theory \emph{Ann. Phys. (N.Y.)} \textbf{64} 311–335

\bibitem{Choi1975}
Choi M D 1975 Completely positive linear maps on complex matrices \emph{Linear Algebra Appl.} \textbf{10} 285–290

\bibitem{Lindblad1976}
Lindblad G 1976 On the generators of quantum dynamical semigroups \emph{Commun. Math. Phys.} \textbf{48} 119–130

\bibitem{GKS1976}
Gorini V, Kossakowski A and Sudarshan E C G 1976 Completely positive dynamical semigroups of $N$-level systems \emph{J. Math. Phys.} \textbf{17} 821–825

\bibitem{Davies1974}
Davies E B 1974 Markovian master equations \emph{Commun. Math. Phys.} \textbf{39} 91–110

\bibitem{Davies1976}
Davies E B 1976 \emph{Quantum Theory of Open Systems} (London: Academic Press)

\bibitem{Spohn1980}
Spohn H 1980 Kinetic equations from Hamiltonian dynamics: Markovian limits \emph{Rev. Mod. Phys.} \textbf{52} 569–615

\bibitem{BreuerPetruccione2002}
Breuer H-P and Petruccione F 2007 \emph{The Theory of Open Quantum Systems} (Oxford: Oxford Univ. Press)

\bibitem{CaldeiraLeggett1983}
Caldeira A O and Leggett A J 1983 Path integral approach to quantum Brownian motion \emph{Physica A} \textbf{121} 587–616

\bibitem{LeggettRMP1987}
Leggett A J, Chakravarty S, Dorsey A T, Fisher M P A, Garg A and Zwerger W 1987 Dynamics of the dissipative two-state system \emph{Rev. Mod. Phys.} \textbf{59} 1–85

\bibitem{JoosZeh1985}
Joos E and Zeh H D 1985 The emergence of classical properties through interaction with the environment \emph{Z. Phys. B} \textbf{59} 223–243

\bibitem{HuPazZhang1992}
Hu B L, Paz J P and Zhang Y 1992 Quantum Brownian motion in a general environment: exact master equation with nonlocal dissipation and colored noise \emph{Phys. Rev. D} \textbf{45} 2843–2861

\bibitem{AharonovAlbertVaidman1988}
Aharonov Y, Albert D Z and Vaidman L 1988 How the result of a measurement of a component of the spin of a spin-$1/2$ particle can turn out to be 100 \emph{Phys. Rev. Lett.} \textbf{60} 1351–1354

\bibitem{DuckSudarshan1989}
Duck I M, Stevenson P M and Sudarshan E C G 1989 The sense in which a “weak measurement’’ of a spin component yields a value 100 \emph{Phys. Rev. D} \textbf{40} 2112

\bibitem{Dressel2014}
Dressel J, Malik M, Miatto F M, Jordan A N and Boyd R W 2014 Colloquium: Understanding quantum weak values: Basics and applications \emph{Rev. Mod. Phys.} \textbf{86} 307–316

\bibitem{Lundeen2011}
Lundeen J S, Sutherland B, Patel A, Stewart C and Bamber C 2011 Direct measurement of the quantum wavefunction \emph{Nature} \textbf{474} 188–191

\bibitem{HostenKwiat2008}
Hosten O and Kwiat P 2008 Observation of the spin Hall effect of light via weak measurements \emph{Science} \textbf{319} 787–790

\bibitem{Dixon2009}
Dixon P B, Starling D J, Jordan A N and Howell J C 2009 Ultrasensitive beam deflection measurement via interferometric weak value amplification \emph{Phys. Rev. Lett.} \textbf{102} 173601

\bibitem{Jordan2014}
Jordan A N, Martínez-Rincón J and Howell J C 2014 Technical advantages for weak-value amplification: When less is more \emph{Phys. Rev. X} \textbf{4} 011031

\bibitem{Belavkin1992}
Belavkin V P 1989 A new wave equation for a continuous nondemolition measurement \emph{Phys. Lett. A} \textbf{140} 355–358

\bibitem{HudsonParthasarathy1984}
Hudson R L and Parthasarathy K R 1984 Quantum Itô’s formula and stochastic evolutions \emph{Commun. Math. Phys.} \textbf{93} 301–323

\bibitem{GardinerCollett1985}
Gardiner C W and Collett M J 1985 Input and output in damped quantum systems: quantum stochastic differential equations and the master equation \emph{Phys. Rev. A} \textbf{31} 3761–3774

\bibitem{JacobsSteck2006}
Jacobs K and Steck D A 2006 A straightforward introduction to continuous quantum measurement \emph{Contemp. Phys.} \textbf{47} 279–303

\bibitem{DalibardCastinMolmer1992}
Dalibard J, Castin Y and Mølmer K 1992 Wave-function approach to dissipative processes in quantum optics \emph{Phys. Rev. Lett.} \textbf{68} 580–583

\bibitem{Carmichael1993}
Carmichael H 1993 \emph{An Open Systems Approach to Quantum Optics} (Berlin: Springer)

\bibitem{WisemanMilburn1993}
Wiseman H M and Milburn G J 1993 Quantum theory of optical feedback via homodyne detection \emph{Phys. Rev. Lett.} \textbf{70} 548–551

\bibitem{Vijay2012}
Vijay R, Slichter D H and Siddiqi I 2011 Observation of quantum jumps in a superconducting artificial atom \emph{Phys. Rev. Lett.} \textbf{106} 110502

\bibitem{Murch2013}
Murch K W, Weber S J, Macklin C and Siddiqi I 2013 Observing single quantum trajectories of a superconducting qubit \emph{Nature} \textbf{502} 211–214

\bibitem{Hatridge2013}
Hatridge M, Shankar S, Mirrahimi M, Schackert F, Geerlings K, Brecht T, Sliwa K, Abdo B, Frunzio L, Schoelkopf R J and Devoret M H 2013 Quantum back-action of an individual variable-strength measurement \emph{Science} \textbf{339} 178–181

\bibitem{Weber2014}
Weber S J, Chantasri A, Dressel J, Jordan A N, Murch K W and Siddiqi I 2014 Mapping the optimal route between two quantum states \emph{Nature} \textbf{511} 570–573

\bibitem{Minev2019}
Minev Z K, Mundhada S O, Shankar S, Reinhold P, Gutiérrez-Jáuregui R, Schoelkopf R J, Mirrahimi M, Carmichael H and Devoret M H 2019 To catch and reverse a quantum jump mid-flight \emph{Nature} \textbf{570} 200–204

\bibitem{Aspelmeyer2014RMP}
Aspelmeyer M, Kippenberg T J and Marquardt F 2014 Cavity optomechanics \emph{Rev. Mod. Phys.} \textbf{86} 1391–1452

\bibitem{Delic2020Science}
Delić U, Reisenbauer M, Dare K, Grass D, Vuletić V, Kiesel N and Aspelmeyer M 2020 Cooling of a levitated nanoparticle to the motional quantum ground state \emph{Science} \textbf{367} 892–895

\bibitem{Magrini2021Nature}
Magrini L, Rosenzweig P, Bach C, Deutschmann-Olek A, Hofer S G, Hong S, Kiesel N, Kugi A and Aspelmeyer M 2021 Real-time optimal quantum control of mechanical motion at room temperature \emph{Nature} \textbf{595} 373–377

\bibitem{Kafri2014}
Kafri D, Taylor J M and Milburn G J 2014 A classical channel model for gravitational decoherence \emph{New J. Phys.} \textbf{16} 065020

\bibitem{Tilloy2018}
Tilloy A 2018 Ghirardi–Rimini–Weber model with massive flashes \emph{Phys. Rev. D} \textbf{97} 021502

\bibitem{Oppenheim2023}
Oppenheim J 2023 A postquantum theory of classical gravity? \emph{Phys. Rev. X} \textbf{13} 041040

\bibitem{Tebbenjohanns2021Nature}
Tebbenjohanns F, Mattana M L, Rossi M, Frimmer M and Novotny L 2021 Quantum control of a nanoparticle optically levitated in cryogenic free space \emph{Nature} \textbf{595} 378–382

\bibitem{Bose2017}
Bose S, Mazumdar A, Morley G W, Ulbricht H, Toroš M, Paternostro M, Geraci A A, Barker P F, Kim M S and Milburn G 2017 Spin entanglement witness for quantum gravity \emph{Phys. Rev. Lett.} \textbf{119} 240401

\bibitem{MarlettoVedral2017}
Marletto C and Vedral V 2017 Gravitationally induced entanglement between two massive particles is sufficient evidence of quantum effects in gravity \emph{Phys. Rev. Lett.} \textbf{119} 240402

\bibitem{GRW1986}
Ghirardi G C, Rimini A and Weber T 1986 Unified dynamics for microscopic and macroscopic systems \emph{Phys. Rev. D} \textbf{34} 470–491

\bibitem{Pearle1989}
Pearle P 1989 Combining stochastic dynamical state-vector reduction with spontaneous localization \emph{Phys. Rev. A} \textbf{39} 2277–2289

\bibitem{GPR1990}
Ghirardi G C, Pearle P and Rimini A 1990 Markov processes in Hilbert space and continuous spontaneous localization of systems of identical particles \emph{Phys. Rev. A} \textbf{42} 78–89

\bibitem{Bassi2013RMP}
Bassi A, Lochan K, Satin S, Singh T P and Ulbricht H 2013 Models of wave-function collapse, underlying theories, and experimental tests \emph{Rev. Mod. Phys.} \textbf{85} 471–527

\bibitem{ArndtHornberger2014}
Arndt M and Hornberger K 2014 Testing the limits of quantum mechanical superpositions \emph{Nat. Phys.} \textbf{10} 271–277

\bibitem{Hornberger2012RMP}
Hornberger K, Gerlich S, Haslinger P, Nimmrichter S and Arndt M 2012 Colloquium: Quantum interference of clusters and molecules \emph{Rev. Mod. Phys.} \textbf{84} 157–173

\bibitem{Gerlich2011}
Gerlich S, Eibenberger S, Tomandl M, Nimmrichter S, Hornberger K, Fagan P J, Tüxen J, Mayor M and Arndt M 2011 Quantum interference of large organic molecules \emph{Nat. Commun.} \textbf{2} 263

\bibitem{Fein2019}
Fein Y Y, Geyer P, Zwick P, Kiałka F, Pedalino S, Mayor M, Gerlich S and Arndt M 2019 Quantum superposition of molecules beyond 25 kDa \emph{Nat. Phys.} \textbf{15} 1242–1245

\bibitem{Bateman2014} Bateman J, Nimmrichter S, Hornberger K and Ulbricht H 2014 Near-field interferometry of a free-falling nanoparticle from a point-like source \emph{Nat. Commun.} \textbf{5} 4788

\bibitem{Carlesso2022NatPhys}
Carlesso M, Donadi S, Ferialdi L, Paternostro M, Ulbricht H and Bassi A 2022 Present status and future challenges of non-interferometric tests of collapse models \emph{Nat. Phys.} \textbf{18} 243–250

\bibitem{Vinante2016}
Vinante A, Bahrami M, Bassi A, Usenko O, Wijts G and Oosterkamp T H 2016 Upper bounds on spontaneous wave-function collapse models using millikelvin-cooled nanocantilevers \emph{Phys. Rev. Lett.} \textbf{116} 090402

\bibitem{Vinante2017}
Vinante A, Mezzena R, Falferi P, Carlesso M and Bassi A 2017 Improved noninterferometric test of collapse models using ultracold cantilevers \emph{Phys. Rev. Lett.} \textbf{119} 110401

\bibitem{Pontin2020}
Pontin A, Bullier N P, Toroš M and Barker P F 2020 Ultranarrow-linewidth levitated nano-oscillator for testing dissipative wave-function collapse \emph{Phys. Rev. Research} \textbf{2} 023349

\bibitem{Fu1997}
Fu Q 1997 Spontaneous radiation of free electrons in a nonrelativistic collapse model \emph{Phys. Rev. A} \textbf{56} 1806–1811

\bibitem{Arnquist2022}
Arnquist I J \emph{et al.} (Majorana Collaboration) 2022 Search for Spontaneous Radiation from Wave Function Collapse in the Majorana Demonstrator \emph{Phys. Rev. Lett.} \textbf{129} 080401

\end{thebibliography}
\end{document}